\documentstyle[12pt]{article}
\begin{document}
\pagestyle{plain}
\newcommand{\be}{\begin{equation}}
\newcommand{\ee}{\end{equation}}
\newcommand{\bea}{\begin{eqnarray}}
\newcommand{\eea}{\end{eqnarray}}
\newcommand{\vp}{\varphi}
\newcommand{\pr}{\prime}
\newcommand{\sech} {{\rm sech}}
\newcommand{\cosech} {{\rm cosech}}
\newcommand{\psib} {\bar{\psi}}
\newcommand{\cosec} {{\rm cosec}}
\def\vs {\vskip .3 true cm}
\centerline { \bf{ The Contribution of Instanton-Antiinstanton Fluctuaions to}}
\centerline { \bf{ the Ground State Energy in Supersymmetric Quantum Mechanical Models}}
\vs
\centerline { \bf{ Avinash Khare}}
\centerline { Institute of Physics, Sachivalaya Marg,}
\centerline { Bhubaneswar 751005, India.}
\centerline { email: khare@iopb.res.in}
\centerline { IP-BBSR/84-20, October 1984, SLAC-PPF T8447}
\vs
{\bf Abstract}

I show that the method of finding if the ground state energy is
nonzero or not by saturating the functional integral directly with
instanton-anti-instanton type fluctuations is not reliable either for
the double or for the triple well potential models in supersymmetric
quantum mechanics.

\vfill
\eject
\vskip .5 true cm
\vfill

In the last few years, supersymmetry (SUSY) has attracted lot of
attention as it appears to cure the gauge hierarchy problem in the
grand unified theories. If SUSY has anything to do with nature, it is
clear that it must be a badly broken symmetry. There are reasons to 
believe that SUSY may be broken dynamically by nonperturbative
fluctuations rather than spontaneously at the tree level. It may
therefore be worthwhile to understand the mechanism responsible for
dynamical SUSY breaking.

Sometime ago, Witten \cite{1} conjectured that instantons may be
responsible for the dynamical SUSY breaking. Since it is not very easy 
to study this question for field theoretic models in 3+1 dimensions,
many workers have studied it in SUSY quantum mechanics. Witten and
specially Salomonson and van Holten \cite{2} have discussed the celebrated
$x^2 - x^4$ double well potential which has one instanton and have
shown that SUSY is indeed dynamically broken in this example. Later
on, Cooper and Freedman \cite{3} have shown that instantons are not necessary
for the SUSY breaking by considering a $x^4$-potential. However, the
real surprise came when Abbott and Zakrzewski \cite{4} 
showed that
instantons are not even sufficient for SUSY breaking. In particular,
they discussed the potential $x^2 - x^4 - x^6$ having symmetric triple 
well \cite{5} and showed that even though there are two instantons in this
model still SUSY is unbroken. This counter example made it clear that
the connection between instantons and dynamical SUSY breaking, if any, 
is not that straight forward \cite{6}.

In a recent paper, Kaul and Mizrachi \cite{7} have reexamined this question 
and offered an explanation as to why SUSY is dynamically broken for
the $x^2-x^4$ potential with double well while it is unbroken for
the triple well case $(x^2-x^4-x^6)$. These authors saturated the
functional integral directly with instanton-antiinstanton type
fluctuations and showed that whereas in the double well case the
ground state energy is raised by these fluctuations, in the triple
well case it is not. According to these authors, their formalism can
be easily generalized to field theoretical models in higher
dimensions. In view of these attractive features, it may be worthwhile 
to critically analyze this approach. In particular, one would like to
know if the approach gives reliable result for any double or triple
well potential model or not. Further, it would appear from the work of 
Kaul and Mizrachi \cite{7} that whereas SUSY would be dynamically broken
for any double well potential, it would remain unbroken for any triple
well case. Is this conjecture really true ? Unfortunately, as I show
below the answer to all these questions is ``no''. In particular, I  
offer concrete examples with double and triple well potentials and
show that in both cases this method leads to wrong conclusion
regarding the question of SUSY breaking. I also show that unbroken or
broken SUSY has nothing to do with the triple or double well structure 
of the potential.

The Minkowaskian action for a supersymmetric classical particle is
given by
\be\label{1}
A_M = \frac {1}{2} \int dt [\dot x^2 -S^2(x)+i\psi^{T} \dot \psi 
 - S'(x)\psi^T\sigma_2\psi] \, ,
\ee
where $S'(x)$ denotes the derivative with respect to its argument x,
$\dot x$ means derivative with respect to time t 
while $\psi$ is a two-component anticommuting variable. Note that this 
action is invariant under the supersymmetric transformation
\be\label{2}
\delta_{\varepsilon} x = \varepsilon^T\sigma_2\psi, \ \ 
\delta_{\varepsilon}\psi = [-i\sigma_2 \dot x-S(x)]\varepsilon, \ \
\delta_{\varepsilon}\psi^T = \varepsilon^T[i\sigma_2\dot x -S(x)] \, .
\ee

Let us first consider the following double well potential
\be\label{3}
V\equiv \frac {1}{2} S^2 = \frac {a\lambda^2}{9m^2}
(x^2-\frac{\alpha^2 m^2}{\lambda})^2(x^2+\frac{\beta^2m^2} {\lambda}) \, ,
\ee
where
\be\label{4}
\alpha^2 = \frac {3}{2a} [ 1+(1-\frac{2a}{3})^{1/2}] \, , 
\ \ \beta^2 = \frac {3}{a} 
[ (1-\frac{2a}{3})^{1/2}-\frac{1}{2}] \, .
\ee
while $a$ is an arbitrary parameter. In order that $V$ has a double well structure it is necessary that $0<a<9/8$. This ensures that $\beta^2$ is always positive. In fact, if one had chosen $a= 9/8$ then $\beta^2 = 0$ and the potential (3) reduces to the symmetric triple well of \cite{7}. For $0 < a < 9/8$, the potential (3) has two classical ground states denoted by |$\pm >$ corresponding to $x = \pm \alpha m/\sqrt{\lambda}$ and $\psi = 0$. Note that for these ground states the classical vacuum energy is zero so that SUSY remains unbroken at the tree level. What about dynamical SUSY breaking because of nonperturbative effects ? Since for this model, asymptotically $ S\sim x^3$ hence a la Witten \cite{1} it follows that SUSY remains unbroken in this model.

Let us now try to follow the arguments of \cite{7} and see if by
saturating the functional integral directly with
instanton-antiinstanton type fluctuations one would arrive at the
same conclusion or not. For the double well potential (3) there is one
instanton and one antiinstanton solution given by \cite{8}
$\psi = 0$ and 
\be\label{5}
 x_I(t-t_1) =\frac{m}{\sqrt{\lambda}}
(\frac{3\varepsilon}{a^2})^{1/2} 
(1+\frac{9\varepsilon}{a^2})^{-1/4}
\frac{\sinh y}
{[1+\frac{1}{3} (1-[1+\frac{9\varepsilon}{a^2}]^{-1/2})
 \sinh^2 y]^{1/2}} \, ,
\ee
\be\label{6}
x_{\overline I} (t-t_1) = - x_I(t-t_1) \, .
\ee
Here
\be\label{7}
\varepsilon = \frac{1}{2} [ (1-a)+(1-\frac{2a}{3})^{3/2} ] \, ,
\ee
while
\be\label{8}
y = (1+\frac{9\varepsilon}{a^2})^{1/4} m(t-t_1) \, .
\ee
The classical action for both these solutions is the same \cite{8}
\be\label{9}
A_0 =\frac{81m^3\varepsilon}{\sqrt 2\lambda a^2}
\frac{F(3,\frac{5}{2}, \frac{11}{4};
  \gamma)\Gamma(11/4)}{(1+\frac{9\varepsilon}{a^2})^{1/4} 
[1-(1+\frac{9\varepsilon}{a^2})^{-1/2}]^3 \Gamma(5/2)\Gamma(1/4)}
\ee
where
\be
\gamma =
-\frac{[2(1+\frac{9\varepsilon}{a^2})^{1/2}+1]}{[(1+\frac{9\varepsilon}{a^2})^{1/2} 
  -1]} \, .
\ee
Now one can essentially run through the arguments of \cite{7}. As is
well known, single instanton or antiinstanton will not contribute to
the vacuum functional integral because of the zero modes of the
relevant fermion determinant obtained by integrating over fermionic
degrees of freedom. However, instanton-antiinstanton configurations
will contribute since there are no exact fermionic zero modes for such 
configurations. By following through the arguments of \cite{7} one can
calculate the lowest order effect from such fluctuations and one finds 
that one has two ground states with equal energies given by
\be\label{11}
< +\mid H\mid +> = <-\mid H\mid ->\simeq \frac{m\hbar}{\sqrt 2\pi} e^{-2A_0/\hbar} \, ,
\ee
where $A_0$ is as given by eq. (9). Thus, according to this approach,
to the lowest order, SUSY is dynamically broken which is contrary to the
exact result that SUSY is unbroken for the potential (3). This counter 
example clearly demonstrates that the method of \cite{7} is not
reliable for double well potentials. Further it also shown that SUSY
need not necessarily be broken for double well potentials.

It may be worthwhile to point out that for the potential (3) if one
had instead followed the approach of \cite{2} and calculated the single 
instanton induced vacuum expectation values of SUSY generators then
also one would have arrived at the wrong conclusion that SUSY is
dynamically  broken for this model.

Let us now turn to the case of the triple well potentials and enquire
about the reliability of the approach proposed in \cite{7}. Let us
consider the following triple well potential
\be\label{12}
V\equiv \frac{1}{2}S^2 =\frac{\lambda^3 x^2}{2m^4}
(x^2-\frac{m^2}{\lambda})^2 (x^2 +\frac{m^2}{\lambda}) \, .
\ee
The potential (12) has three classical ground states denoted by 
$\mid +>, \ 
\mid ->$ and $\mid 0 >$ corresponding to $x = + m/\sqrt{\lambda}, 
 - \frac{m}{\sqrt\lambda}$, 0 respectively (and $\psi = 0$). Note that the vacuum 
 energy is zero for all these ground states so that SUSY is unbroken at 
 the tree level. What about dynamical SUSY breaking ? Since
 asymptotically $s\sim x^4$ hence a 1a Witten  \cite{1} it follows that the 
 ground state energy $E_0 > 0$ and hence SUSY is dynamically broken in 
 this model.

Let us now follow the approach of \cite{7} and see if by saturating the 
functional integral directly with non-perturbative fluctuations one
arrives at the same conclusion or not. For the triple well potential
(12) it is clear that there will be two types of instantons which will
interpolate between the classical ground states $x = 0$ 
and $x = \frac
{m}{\sqrt\lambda}$ as well as $x= - \frac{m}{\sqrt\lambda}$ and 
 $x = 0$ as 
$t$ goes from 
$\rightarrow -\infty$ to $t\rightarrow \infty$ respectively. Similarly
there will be two types of antiinstantons which will interpolate
between the classical ground states $x = \frac{m}{\sqrt\lambda}$
and $x =
0$ as well as $x = 0$ and  $x = - m/\sqrt\lambda$ as 
$t$ goes from $\rightarrow -\infty$ to 
$t\rightarrow \infty$ respectively.
The action for all these instantons and antiinstantons is the same:
\be\label{13}
A_0 = \frac{4m^3}{15\lambda} \, .
\ee
Single instanton or antiinstanton again does not contribute to the
vacuum functional integral because of the fermionic zero modes. The
possible next order contribution to the vacuum functional integral
will again come from instanton-antiinstanton configurations. In this
case, there will also be contributions from two instantons and two
antiinstantons representing quantum tunneling from $\mid - >$ to
$\mid + >$ and $\mid +>$ to $\mid ->$. One can calculate the lowest order
effect from such vacuum fluctuations by following the approach of
\cite{7}. One would then find that to the lowest order
\be\label{14}
< 0\mid H\mid 0 > _{x_I^{(+)}x_{\overline I}^{(+)}} = 
< 0\mid H\mid 0 > _{x_I^{(-)}x_{\overline I}^{(-)}} \simeq
\frac{m\hbar}{\sqrt 2\pi} e^{-2A_0/\hbar} = X \, ,
\ee
so that
\be\label{15}
< 0\mid H\mid 0 > _{x_I^{(+)}x_{\overline I}^{(+)} 
+x_I^{(-)}x_{\overline I}^{(-)}} = 2X \, .
\ee
Further
\bea\label{16}
&& < +\mid H\mid + > _{x_I^{(+)}x_{\overline I}^{(+)}} = 
< -\mid H\mid - > _{x_I^{(-)}x_{\overline I}^{(-)}} 
=< +\mid H\mid - > _{x_I^{(+)}x_{I}^{(-)}} \nonumber \\ 
&& = < -\mid H\mid + > _{x_{\overline I}^{(+)}x_{\overline I}^{(-)}} 
\simeq
\frac{m\hbar}{\sqrt 2\pi} e^{-2A_0/\hbar} = X \, , 
\eea 
to the same order. Here $A_0$ is as given by eq.(13). Hence to the
lowest order, the Hamiltonian matrix for the low-lying states can be
written as
\be\label{17} 
H = \left ( \begin{array}{ccc} x & 0 & x \\
0 & 2x & 0 \\ x & 0 & x  \end{array} \right ) \, .
\ee 
The matrix has a zero eigenvalue and two equal non-zero
eigenvalues. Thus according to the approach of \cite{7}, 
to the lowest order, 
SUSY is not dynamically broken in this model which is contrary to the
exact result. Thus the approach of \cite{7} does not give reliable
answers for the triple well potentials either. Further, this counter
example also demonstrates that SUSY need not necessarily be unbroken
for triple well potentials.

Summarizing, it appears that the approach of \cite{7} at least in its
present form, does not give reliable answers for either the 
double or the triple
well potentials. Perhaps, it is not enough to calculate the
contributions from widely separated instantons and antiinstanton but
one must also include the contribution from their mutual
interactions. Clearly much more work is necessary in order to clarify
the connection between instantons and dynamical SUSY breaking.

\end{document}